\documentstyle[prl,aps,epsf,floats]{revtex}
\begin{document}
\draft
\twocolumn[\hsize\textwidth\columnwidth\hsize\csname @twocolumnfalse\endcsname
\title { Giant enhancement of anisotropy  
         by electron-phonon interaction } 

\author{P.\,E.\,Kornilovitch}
\address{Blackett Laboratory, Imperial College,
Prince Consort Road, London SW7 2BZ, United Kingdom}

\date{\today}
\maketitle

\begin{abstract}

Anisotropic electron-phonon interaction is shown to lead to the anisotropic 
polaron effect. The resulting anisotropy of the polaron band is an 
exponential function of the electron-phonon coupling and might
be as big as $10^3$. This also makes anisotropy very sensitive
to small changes of coupling and implies wide variations of anisotropy 
among compounds of similar structure. The isotope effect on mass 
anisotropy is predicted. Polaron masses  are obtained by an exact 
Quantum Monte Carlo method. Implications for high-temperature 
superconductors are briefly discussed.

\end{abstract}
\vskip2pc]
\narrowtext

In an anisotropic crystal electron-phonon interaction (EPI) is
anisotropic too. This is of minor importance in a metal with a
high carrier density and weak EPI. The carrier mass is renormalized as 
$m^{\ast}_{\alpha} = m_{0\alpha} (1 + \gamma_{\alpha} \lambda)$,
where $\lambda$ is the electron-phonon coupling constant and
$\gamma_{\alpha} \sim 1$ a numerical coefficient corresponding to
direction $\alpha$. At $\lambda \ll 1$ mass renormalization 
is small and EPI has little effect on the band anisotropy which is 
governed by bare anisotropy of the rigid lattice.
The situation is qualitatively different in a semiconductor
with a low carrier density and strong EPI $\lambda \geq 1$. Under 
these conditions formation of small (lattice) polarons is
expected \cite{Appel,Firsov,Alexandrov&Mott}. A general property
of {\em small} polarons is an exponential renormalization of mass
$m^{\ast}_{\alpha} \propto \exp (\gamma_{\alpha}\lambda/\bar\omega)$,
where $\bar\omega$ is the dimensionless phonon frequency \cite{Tjablikov}. 
Now the mass anisotropy $m^{\ast}_{\alpha}/m^{\ast}_{\beta} \propto
\exp ( (\gamma_{\alpha}$$-$$\gamma_{\beta}) \lambda/ \bar\omega)$  
is an {\em exponential} function of $\lambda$, $\bar\omega$, and
the form of EPI (through $\gamma_{\alpha}$s). This simple relation
hints several possible effects. (i) Anisotropy may be very large if
$\lambda \geq 1$ and $\bar\omega \leq 1$.
(ii) Anisotropy is very sensitive to small changes of $\lambda$
caused, e.g., by doping (due to changes in screening). This also
implies that different compounds that are similar in structure and
have close $\lambda$ may nevertheless have very different anisotropies.
(iii) The dependence on phonon frequency is strong, hence
the {\em isotope effect} on anisotropy. Thus anisotropic
EPI may have profound effect on electronic properties of
doped semiconductors.

The anisotropic case has been little touched in the polaron literature.
Kahn considered the large (continuum) Fr\"ohlich polaron with bare 
anisotropy and isotropic EPI and found that the polaron is {\em less} 
anisotropic than the bare carrier \cite{Kahn}. Recently Caprara and 
Del Prete found an {\em enhancement} of anisotropy by a factor 
$\approx$$2$ in the vicinity of a saddle point of the bare spectrum 
\cite{Caprara}. Both studies were weak-coupling perturbational. 
In the strong-coupling regime the Lang-Firsov transformation 
\cite{Lang} leads quite naturally to a strongly anisotropic polaron 
but we are not aware of any specific applications of this method. 
Also we are not aware of any previous numerical studies of the 
anisotropic case. Almost all the recent activity in the polaron 
field has been directed to the paradigmatic Holstein model 
\cite{Holstein} where EPI is local and the effect is absent. 

The recently developed Quantum Monte Carlo (QMC) algorithms
\cite{Prokof'ev_two,Kornilovitch} have opened the 
possibility to calculate the polaron effective mass exactly. 
The methods also allow studies of infinite 
lattices and arbitrary forms of electron-phonon interaction 
including anisotropic. In this paper we apply the method of 
Ref. \cite{Kornilovitch}, based on Feynman's integration of phonons 
\cite{Feynman,DeRaedt}, to two models with anisotropic EPI. The first 
one is a simple two-dimensional (2D) model where the enhancement 
of anisotropy is weak. The second model is a three-dimensional (3D) 
one with long-range EPI. Here we will find a very large enhancement 
and confirm the existence of the effects mentioned in the beginning.    

\begin{figure}[t]
\begin{center}
\leavevmode
\hbox{
\epsfxsize=8.4cm
\epsffile{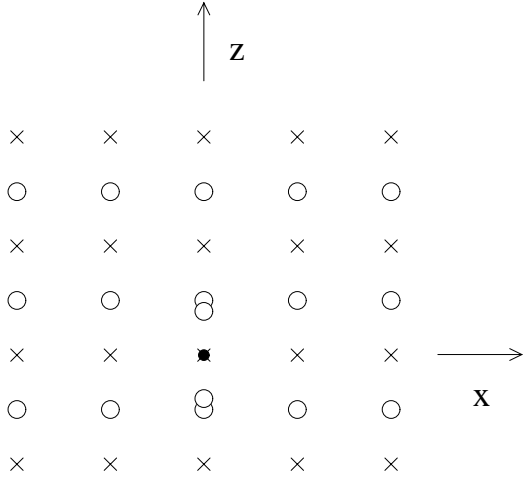}
}
\end{center}
\vspace{-0.5cm}
\caption{
The two-dimensional lattice (positioned in the plane $xz$)
of the model (\ref{one}). The carrier (bullet) hops between {\bf n} 
sites (crosses) but interacts with {\bf m} sites (circles) which 
vibrate along $z$-direction. Two {\bf m} sites are shifted from 
their equilibrium positions by some distance forming a small polaron. 
The three-dimensional lattice of the model (\ref{three}) is obtained 
by repeating the whole figure along $y$-direction perpendicular to 
the sheet plane.  
}
\label{fig1}
\end{figure}

\begin{figure}[t]
\begin{center}
\leavevmode
\hbox{
\epsfxsize=8.4cm
\epsffile{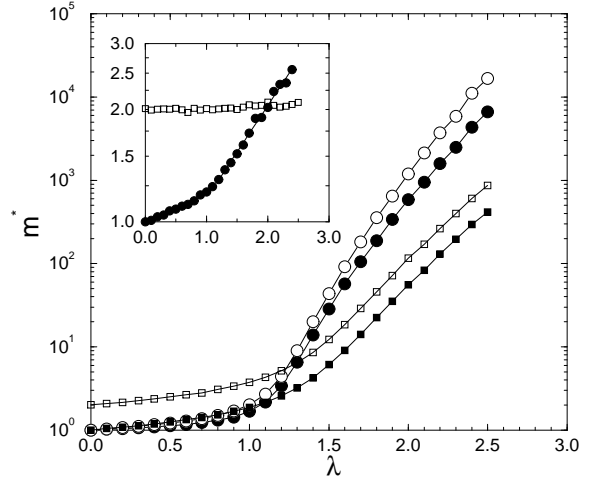}
}
\end{center}
\vspace{-0.5cm}
\caption{ 
Polaron masses in the model (\ref{one}) for $t_{\perp} = t$ (circles) 
and in the 2D Holstein model for $t_{\perp}/t = 0.5$ (squares). 
$\bar\omega = 1.0 $ in both cases. Open (filled) symbols are 
$m^{\ast}_z$ ($m^{\ast}_x$). Inset: anisotropy $m^{\ast}_z/m^{\ast}_x$ 
for the model (\ref{one}) (circles) and Holstein model (squares). 
}
\label{fig2}
\end{figure}

Let us begin with the simplest possible 2D model with anisotropic
EPI (see Fig.~\ref{fig1}). The lattice consists of 
two interpenetrating square sublattices ${\bf n}$ (crosses) and
${\bf m}$ (circles). The sites ${\bf n}$ are fixed in their positions
but sites ${\bf m}$ can vibrate along $z$-direction (up-down),
$\xi_{\bf m}$ being the internal coordinates. All ${\bf m}$ are 
uncoupled and have the same frequency $\omega$ and mass $M$.
The carrier moves in the sublattice ${\bf n}$ by hopping 
between nearest neighbors with amplitudes $t_{\bf nn'}$.
The carrier at a site ${\bf n}$ attracts the two nearest
sites ${\bf m}$ (above and below) with {\em force} $\kappa$. 
The model Hamiltonian is
\begin{eqnarray} 
H = & - & \sum_{\bf \langle nn' \rangle} t_{\bf nn'} 
c^{\dagger}_{\bf n} c_{\bf n'} +
\sum_{\bf m} \left( \frac{\hat p^2_{\bf m}}{2M} + 
\frac{M\omega^2}{2} \xi^2_{\bf m} \right) \nonumber \\ 
    & - & \sum_{\bf \langle nm \rangle} f_{\bf m}({\bf n})
c^{\dagger}_{\bf n} c_{\bf n} \xi_{\bf m} ,
\label{zero}
\end{eqnarray}
\begin{equation}
 f_{\bf m}({\bf n}) = \kappa 
( \delta_{{\bf n}, {\bf m} - {\bf z}/2} - 
\delta_{{\bf n}, {\bf m} + {\bf z}/2})  ,
\label{one}
\end{equation}
where $\hat p_{\bf m} = -i\hbar \partial/\partial \xi_{\bf m}$. The 
hopping integral in $x$ and $z$ directions are $t$ and $t_{\perp}$, 
respectively. The model is characterised by the dimensionless
frequency $\bar\omega =\hbar \omega/t$, the bare anisotropy
ratio $m_{0z}/m_{0x} = t/t_{\perp}$, and the coupling constant
$\lambda \equiv  [\sum_{\bf m} f^2_{\bf m}(0)]/(2M\omega^2D) = 
2 \kappa^2/(2M\omega^2D)$, where $D$ is half of the bare bandwidth.
If the carrier interacted only with {\em one} neighboring ${\bf m}$
site then Eq.~(\ref{one}) would have described the 2D Holstein model. 
Polaron masses of the latter model for $\bar \omega$$=$$1.0$
and $t_{\perp} = 0.5 \, t$, obtained with QMC, are shown in 
Fig.~\ref{fig2} (main picture, squares). 
At $\lambda$$>$$1.5$ both $m^{\ast}_z$ and $m^{\ast}_x$ increase 
expo\-nen\-ti\-al\-ly as $\propto$$\,\exp(4.2\, \lambda)$ .
However, the ratio $m^{\ast}_z/m^{\ast}_x$ re\-mains constant as a
function of $\lambda$ and equal to the bare anisotropy 2 
(see inset, squares). Thus the local interaction of the 
Holstein model does not enhance bare anisotropy.

The behavior of the model (\ref{one}) is different. While the
renormalization of both masses is similar to the Holstein case
$m^{\ast}_z$ now grows faster than $m^{\ast}_x$ even for 
$t_{\perp} = t$. The ratio $m^{\ast}_z/m^{\ast}_x$ grows
exponentially as $\propto \exp(0.58 \, \lambda)$  
in the small-polaron regime $\lambda > 1.2$ (see the inset in 
Fig. \ref{fig2}, circles). The physical reason for the 
difference is as follows. In the small-polaron regime the carrier 
is mostly confined to a single lattice site {\bf n} and both 
nearest {\bf m} ions are shifted by some distance $d$ from their 
equilibrium positions (see Fig.~\ref{fig1}). Tunneling
along $x$-direction occurs when two neighboring {\bf m} ions    
on the right (or left) get shifted fluctuationally by $d$
which creates a potential well on the neighboring {\bf n}
site, similar to the one the electron is already in. However,
for the tunneling along $z$-direction to become possible only one 
{\bf m} ion must be shifted by $d$ but another, the one which
is forming the current potential well, must be shifted by the
{\em double} distance $2d$. This process is less probable and
requires more time to occur. That is why, although the lattice
deformation extends in $z$-direction, it is more mobile in 
$x$-direction.

The simple model considered illustrates the two generic
properties of strong anisotropic electron-phonon interaction:
(i) it makes the spectrum anisotropic, and
(ii) anisotropy is an exponential function of coupling.
However, the actual enhancement is modest in this model,
$m^{\ast}_z/m^{\ast}_x \approx 2$ for $\lambda = 2.0$ when
the masses themselves are already huge, $m^{\ast} \sim 10^3$. In order
to understand under which conditions anisotropy can be larger
one has to consider how the form of electron-phonon interaction, 
i.e. the shape of the function $f_{\bf m}({\bf n})$ affects
mass enhancement of the small polaron. Within the path-integral
approach the inverse polaron mass is the second moment of the end-to-end 
distribution of imaginary-time paths with open boundary conditions 
\cite{Kornilovitch}. Each path includes in its statistical weight 
a phonon-induced factor $e^{A}$ where $A$ is the action of the 
retarded self-interaction \cite{Feynman}. Essentially, $A$ depends 
on the forces $f_{\bf m}({\bf n})$ via combinations
\begin{equation}
\Phi({\bf r}_1 - {\bf r}_2) = \sum_{\bf m} 
f_{\bf m}({\bf r}_1) f_{\bf m}({\bf r}_2) .
\label{two}
\end{equation}
Here {\bf r} is the carrier position on a path, i.e., one of {\bf n}.
Normally, the function $\Phi({\bf r})$ is the largest at ${\bf r}=0$
and decays monotonically at large ${\bf r}$. What matters for polaron
mass are the {\em gradients} $\nabla \Phi({\bf r})$ in the 
corresponding direction. The smaller the gradient the less the 
action changes upon the increase of ${\bf r}_1 - {\bf r}_2$,
the more likely such a change will be accepted by the Monte Carlo
process, the larger the second moment of the resulting end-to-end
distribution. Thus, a steep (smooth) $\Phi({\bf r})$ leads to a large
(small) effective mass. 

\begin{figure}[t]
\begin{center}
\leavevmode
\hbox{
\epsfxsize=8.4cm
\epsffile{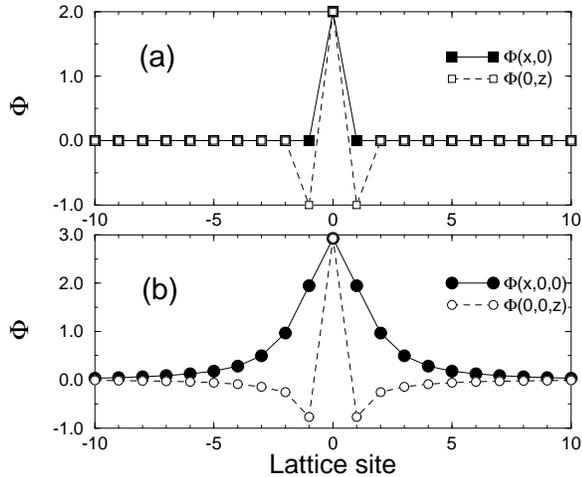}
}
\end{center}
\vspace{-0.5cm}
\caption{
Function $\kappa^{-2} \Phi({\bf r})$ [Eq.~(\ref{two})] for: 
(a) - the 2D model (\ref{one}); (b) - the 3D model (\ref{three}).  
}
\label{fig3}
\end{figure}

For the Holstein model 
(for which $f_{\bf m}({\bf n})$$=$$\kappa \delta_{\bf mn}$) 
function (\ref{two}) is the lattice delta function 
$\Phi({\bf r}_1$$-$${\bf r}_2) = \kappa^2 \delta_{{\bf r}_1 {\bf r}_2}$ 
which is a steep function. Therefore in the small-polaron
regime the weight of paths with non-zero ${\bf r}_1 - {\bf r}_2$
is exponentially small leading to an exponentially small second moment.
For the model (\ref{one}) the $x$- and $z$- cross-sections of 
$\Phi({\bf r})$ are shown in Fig. \ref{fig3}(a). One can see that due to
negative values at $(x=0, z=\pm 1)$ the gradient in $z$ direction 
is bigger than in $x$ direction. This is the reason for the 
enhanced $m^{\ast}_z$.
There are cases where $\nabla \Phi({\bf r})$ in one 
direction is much smaller than in the Holstein model or than in
model (\ref{one}). Consider again a lattice with the topology of
Fig. \ref{fig1}. Elementary geometrical considerations
show that $\nabla_x \Phi({\bf r})$ is small when (i) the 
distance between chains {\bf n} and {\bf m} is large; 
(ii) interaction with {\em distant} {\bf m} ions is included.
When these conditions are fulfilled one should
expect a much lighter $x$-mass than in the Holstein case. 
At the same time $\nabla_z \Phi({\bf r})$ is {\em always} big
at the origin. Indeed, by definition $\Phi(0,0)$ is positive
but $\Phi(0,\pm 1)$ is negative (because $f_{\bf m}(0,0)$
and $f_{\bf m}(0,1)$ have different signs for {\bf m} ions with
${\bf m} = (x, 1/2)$). Therefore the $z$-mass is of the order
of the Holstein mass and, consequently, a large anisotropy 
$m^{\ast}_z/m^{\ast}_x$ might follow. One should add, that
if {\bf m} ions were allowed to move along $x$-direction
it would have increased $m^{\ast}_x$ and decreased $m^{\ast}_z$.
A predominant interaction with a single phonon polarization is 
essential for a large anisotropy. Finally, it is not difficult to 
see that in 3D the difference between $\nabla_z \Phi({\bf r})$
and $\nabla_x \Phi({\bf r})$ would be even bigger than in 2D.

We now consider a 3D model where all the above factors are present.
The lattice structure is obtained from the one depicted in 
Fig.~\ref{fig2}, by repeating it along $y$-direction which is
perpendicular to the sheet plane. The distance between neighboring 
{\bf n} and {\bf m} {\em planes} is equal to the lattice constant 
in $x$ and $y$ directions (i.e., the lattice constant in $z$ direction
is {\em twice} the one in $xy$ plane). The EPI has the form
\begin{equation}
f_{\bf m}({\bf n}) = \kappa \frac{m_z - n_z}{| {\bf m} - {\bf n} |^{3/2}}
\, e^{- \frac{| {\bf m} - {\bf n} |}{R} }  , 
\label{three}
\end{equation}
which is the $z$-projection of the Coulomb force with a screening
radius $R$. This model is designed to describe the layered structure
of high-$T_c$ superconductors, {\bf n} sites (crosses) representing
the copper-oxygen planes and {\bf m} sites (circles) the apical oxygens
\cite{Alexandrov}. Its 1D (one {\bf n} chain and one {\bf m} chain) 
and 2D (one {\bf n} plane and one {\bf m} plane) versions were considered 
in \cite{Alexandrov&Kornilovitch} (see also \cite{Kornilovitch}). 
There was found a much reduced polaron mass, as compared with the 
Holstein model. Here we consider the full 3D problem. 
Due to the low $c$-axis conductivity of the cuprates 
screening in $z$ direction is poor and $R$ is large.
We use $R = 10$ lattice constants in $x$ direction. For this
interaction $\lambda = 2.93 \kappa^2/(2 M \omega^2 D) $. Function
$\Phi({\bf r})$ for this model is shown in Fig. \ref{fig3}(b).

\begin{figure}[t]
\begin{center}
\leavevmode
\hbox{
\epsfxsize=8.4cm
\epsffile{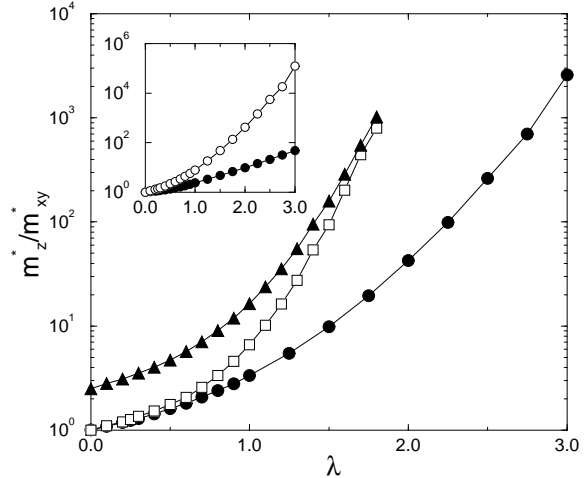}
}
\end{center}
\vspace{-0.5cm}
\caption{ 
Mass anisotropy of the model (\ref{three}). 
Circles - $\bar\omega = 1.0$, $t_{\perp} = 0.25\, t$;
squares - $\bar\omega = 0.5$, $t_{\perp} = 0.25\, t$;
triangles - $\bar\omega = 0.5$, $t_{\perp} = 0.1\, t$.
Inset: masses $m^{\ast}_{xy}$ (filled circles) and $m^{\ast}_z$ for
$\bar\omega = 1.0 $, $t_{\perp} = 0.25\, t$.
}
\label{fig4}
\end{figure}

Effective masses of model (\ref{three}) for $\bar\omega = 1.0$
and $t_{\perp} = 0.25 \,t$, obtained with the exact QMC, are shown 
in the inset of Fig \ref{fig4}. There is a striking difference in 
$\lambda$-dependences of $m^{\ast}_{xy}$ and $m^{\ast}_z$.
While $m^{\ast}_{xy}$ grows much slower than the Holstein mass,
which is consistent with previous findings in one and two dimensions, 
$m^{\ast}_z$ grows similar to the Holstein case. For example,
for $\lambda=2.0$ $m^{\ast}_z = 420 \, m_{0x}$ while 
$m^{\ast}_x = 10\, m_{0x}$, for $\lambda=2.5$ 
$m^{\ast}_z = 5600 \, m_{0x}$ while $m^{\ast}_x = 21\, m_{0x}$, etc.  
As a result, the mass anisotropy increases sharply with coupling
reaching 2500 for $\lambda=3.0$ (see the main picture in 
Fig.~\ref{fig4}, circles). Basically, we are dealing with a new 
situation when two types of polaron are present at the same time.
In $z$ direction the carrier is the Holstein small polaron with a small
size of lattice deformation (of the order of one lattice constant),
localization of the carrier within this local deformation, and
a very large effective mass. In $xy$ plane the lattice deformation 
is larger than both the lattice constant and the localization
area of the carrier, and the polaron is much more mobile 
(such a quasiparticle was named small Fr\"ohlich polaron in 
\cite{Alexandrov&Kornilovitch}). It is interesting to note fits 
$m^{\ast}_z/m_{0x} = \exp (1.26 \lambda + 0.88 \lambda^2)$ and
$m^{\ast}_z/m^{\ast}_{xy} = \exp ( 0.42 \lambda + 0.71 \lambda^2)$.
The quadratic terms in the exponents are unusual. We were
unable to establish if this is the true asymptotic behavior or
the dependencies approach a pure exponential growth at larger 
$\lambda$.
 
We continue to study the model (\ref{three}) by changing its parameters.
A general property of the small polaron is that its mass increases
as the phonon frequency decreases because the lattice deformation
becomes less mobile, $m^{\ast}_{\alpha} \propto 
\exp( \gamma_{\alpha} \lambda/\bar\omega )$. In our case 
$\gamma_z > \gamma_{xy}$. Therefore one should expect much stronger 
change of $m^{\ast}_z$ than of $m^{\ast}_{xy}$ with frequency. 
Figure~\ref{fig4} shows $m^{\ast}_z/m^{\ast}_{xy}$ 
for $\bar\omega = 0.5$ and $t_{\perp} = 0.25\, t$ (squares).    
One can see that in this case the anisotropy is much bigger than 
when $\bar\omega = 1.0$. This implies the existence of the
{\em isotope effect on mass anisotropy} in systems with
strong long-range anisotropic electron-phonon interaction.
Another model parameter is the bare anisotropy $t_{\perp}/t$.
It has been found to be of minor importance. In short, it is a 
constant factor which is carried throughout the whole $\lambda$ 
interval. Figure~\ref{fig4} shows $m^{\ast}_z/m^{\ast}_{xy}$ for 
$\bar\omega = 0.5$ and $t_{\perp} = 0.1\, t$ which corresponds 
to the bare anisotropy $m_{0z}/m_{0xy} = 2.5$ (triangles). 
The plot is very close to the one with $t_{\perp} = 0.25 \, t$, 
the small difference resulting from the latter case 
being more adiabatic than $t_{\perp} = 0.1\, t$.  

Layered high-$T_c$ superconductors seem to be a good candidate
for the effect described. Indeed, in the cuprates the carrier
density is low, EPI is obviously anisotropic and {\em strong},
as revealed by isotope substitution \cite{Franck,Zhao} and
neutron scattering \cite{Egami} experiments. At the same time, 
a number of unusual $c$-axis properties \cite{Cooper} 
--- anomalously low interlayer hopping (large $m^{\ast}_z$),
anomalously large  mass anisotropy, large variation of anisotropy
among different cuprates --- are reproduced in our model
calculation. Although other effects, such as the pairing of 
carriers, have been shown to account for some $c$-axis anomalies 
of the cuprates \cite{Alex&Kabanov&Mott}, the effect of strong 
anisotropic EPI seems to be worth further studying. For instance, 
the puzzling exponential increase of $c$-axis
plasma frequency with doping and the exponential decrease
of anisotropy with doping observed in YBCO and other cuprates
\cite{Cooper} can be explained easily by the present mechanism if 
one assumes that extra holes screen the long-range EPI and 
thereby reduce $\lambda$ \cite{Alex&Kabanov&Mott}.

In summary, it has been argued that strong
anisotropic electron-phonon interaction in general  
enhances the anisotropy of the carrier band. The effect
is strongest when (i) the interaction is predominantly with
one phonon polarization only; (ii) the interaction is long-range; 
(iii) the dimensionality is three; (iv) the distance
between planes is large. The mass anisotropy is an exponential
function of coupling which results in the large value of anisotropy 
and in wide variations among different compounds of
similar structure. The bare anisotropy is irrelevant for the
effect. With decreasing phonon frequency the mass anisotropy 
increases drastically. This amounts to isotope shift of the mass
anisotropy. Possible candidates for the effect are layered 
high-$T_c$ superconductors. 

The author is grateful to A.\,S.\,Alexandrov, D.\,M.\,Eagles,
W.\,M.\,C.\,Foulkes, A.\,O.\,Gogolin, and V.\,V.\,Kabanov for useful
discussions. This work was supported by EPSRC grant GR/L40113.

\end{document}